\documentclass[aps,prl,preprint,superscriptaddress]{revtex4}

\usepackage{graphicx}
\begin{document}

\title{Gradual Disappearance of the Fermi Surface near the Metal-Insulator Transition
 in La$_{1-x}$Sr$_{x}$MnO$_{3}$}

\author{A. Chikamatsu}
\affiliation{Department of Applied Chemistry, The University of Tokyo, 7-3-1 Hongo, Bunkyo-ku, Tokyo 113-8656, Japan}
\author{H. Wadati}
\affiliation{Department of Physics, The University of Tokyo, 7-3-1 Hongo, Bunkyo-ku, Tokyo 113-8656, Japan}
\author{H. Kumigashira}
\email[Author to whom correspondence should be addressed; Electronic mail: ]{kumigashira@sr.t.u-tokyo.ac.jp}
\affiliation{Department of Applied Chemistry, The University of Tokyo, 7-3-1 Hongo, Bunkyo-ku, Tokyo 113-8656, Japan}
\author{M. Oshima}
\affiliation{Department of Applied Chemistry, The University of Tokyo, 7-3-1 Hongo, Bunkyo-ku, Tokyo 113-8656, Japan}
\author{A. Fujimori}
\affiliation{Department of Physics, The University of Tokyo, 7-3-1 Hongo, Bunkyo-ku, Tokyo 113-8656, Japan}
\affiliation{Department of Complexity Science and Engineering, University of Tokyo, Kashiwa 277-8561, Japan}
\author{M. Lippmaa}
\affiliation{Institute for Solid State Physics, The University of Tokyo, Kashiwa 277-8581, Japan}
\author{K. Ono}
\affiliation{Institute of Materials Structure Science, High Energy Accelerator Research Organization, Tsukuba 305-0801, Japan}
\author{M. Kawasaki}
\affiliation{Institute for Materials Research, Tohoku University, Sendai 980-8577, Japan}
\author{H. Koinuma}
\affiliation{Department of Complexity Science and Engineering, University of Tokyo, Kashiwa 277-8561, Japan}

\date{\today}

\begin{abstract}
We report the first observation of changes in the electronic structure of La$_{1-x}$Sr$_{x}$MnO$_{3}$ (LSMO) across the filling-control metal-insulator (MI) transition by means of \textit{in situ} angle-resolved photoemission spectroscopy (ARPES) of epitaxial thin films.  The Fermi surface gradually disappears near the MI transition by transferring the spectral weight from the coherent band near the Fermi level ($E_{F}$) to the lower Hubbard band, whereas a pseudogap behavior also exists in the ARPES spectra in the close vicinity of $E_{F}$ for the metallic LSMO.  These results indicate that the spectral weight transfer derived from strong electron-electron interaction dominates the gap formation in LSMO associated with the filling-control MI transition.
\end{abstract}


\maketitle

Change in band filling or carrier-doping procedure occasionally drives the metal-insulator (MI) transition in strongly correlated electron systems like perovskite-type transition metal oxides \cite{ImadaM:1998}.  Among them, the carrier-doped perovskite-type manganese oxides exhibit a rich phase diagram originating from the competition among different electronic phases due to the close interplay of spin, charge, orbital, and lattice degrees of freedom \cite{ImadaM:1998, UrushibaraA:1995, HembergerJ:2002, TomiokaY:1996}.  The key issue in elucidating the mechanism of the MI transition is to understand how the electronic structure evolves with hole doping from the Mott insulator to the ferromagnetic metallic phase.  The hole-doped La$_{1-x}$Sr$_{x}$MnO$_{3}$ (LSMO) is an ideal system to address the doping evolution of the electronic structure because of its relatively large bandwidth and the absence of long-range orbital and/or charge ordering as observed in Pr$_{1-x}$Ca$_{x}$MnO$_{3}$ \cite{TomiokaY:1996}.  The parent compound LaMnO$_{3}$ is an antiferromagnetic insulator, while hole-doping produces a ferromagnetic insulating phase (FI) in a fairly narrow $x$ region around $x = 0.1$ and then ferromagnetic metallic phases (FM) for $0.16 < x < 0.5$ \cite{UrushibaraA:1995}.  In the ferromagnetic metallic phase, LSMO shows colossal magnetoresistance (CMR), associated with a temperature-induced MI transition and half-metallic conductivity \cite{ImadaM:1998, ParkJH:1998}.  In order to clarify the doping-induced changes of the electronic properties, one has to understand the changes that occur in the band structure, especially in the Fermi surface (FS), across the MI transition.

Angle-resolved photoemission spectroscopy (ARPES), by which one can directly determine the band structure and the FS, is an ideal tool for resolving such an issue \cite{ShenZX:1995}. Recently, the (remnant) FS topology and quasiparticle (QP) states coupled to a collective excitation in manganites have been intensively investigated by ARPES measurements \cite{SunZ:2006, MannellaN:2005, ShiM:2004, ChikamatsuA:2006}: ARPES studies for layered manganites \cite{SunZ:2006, MannellaN:2005} have revealed the importance of electron-phonon interaction for (pseudo)gap formation in manganites and the existence of QP and a pseudogap state with nodal-antinodal dichotomous character similar to the characteristic feature of the copper oxide high-temperature superconductors \cite{MannellaN:2005}.  Relevant many-body interactions in strongly correlated oxides are either electron-electron interaction or electron-photon interaction and these interactions are expected to be strong in manganites.  Hence it is essential to clarify how the relevant interactions play roles in the filling-control MI transition of manganites.  However, the lack of information about the gap or pseudogap formation in the manganites across their filling-control MI transition has largely obscured the understanding for the novel electronic properties of the manganites.

In this Letter, we report the first observation for the carrier-doping evolution of the band structure near $E_{F}$ in LSMO thin films, which provides a comprehensive understanding the gap or pseudogap formation in manganites associated with the filling-control MI transition.  The conduction band, which forms an electron FS at the $\Gamma$ point in half-metallic LSMO $x = 0.4$ films, gradually becomes weak with decreasing hole concentration $x$, and almost disappears in the insulating LSMO $x = 0.1$ phase.  Besides the loss of spectral weight near $E_{F}$, the pseudogap behavior also exists in the ARPES spectra in the close vicinity of $E_{F}$ for the metallic LSMO irrespective of $x$.  Comparing the present ARPES results with previous ARPES results as well as the other experiments, we discuss the origin of the novel electronic structure in manganese oxides.

The ARPES measurements were carried out using a photoemission spectroscopy (PES) system combined with a laser MBE chamber, which was installed at beamline BL-1C of the Photon Factory, KEK \cite{ChikamatsuA:2006, HoribaK:2005}.  Approximately 400 {\AA}-thick LSMO films were grown epitaxially on Nb-doped SrTiO$_{3}$ (STO) substrates by pulsed laser deposition.  Details are described in Refs. \cite{ChikamatsuA:2006, HoribaK:2005}.  After deposition, the films were moved into the photoemission chamber under a vacuum of 10$^{-10}$ Torr.  The cleanness of the vacuum-transferred LSMO films was also confirmed by low energy electron diffraction.  The ARPES spectra were taken at 25 K for ferromagnetic metallic LSMO ($x = 0.2, 0.3,$ and 0.4) films using a GAMMADATA SCIENTA SES-100 electron-energy analyzer in the angle-resolved mode, while they were taken at 150 K for a ferromagnetic insulator LSMO $x = 0.1$ to prevent the charge-up effects in ARPES measurements.  The total energy and angular resolution were set at about 150 meV and 0.5$^{\circ}$, respectively.  The Fermi level of the samples was referred to that of a gold foil which was in electric contact with the sample.  Atomic force microscopy, X-ray diffraction, and magnetization measurements were performed for characterization, and results were nearly the same as the reported results \cite{HoribaK:2005}.

\begin{figure}
\begin{center}
\includegraphics[width=14cm]{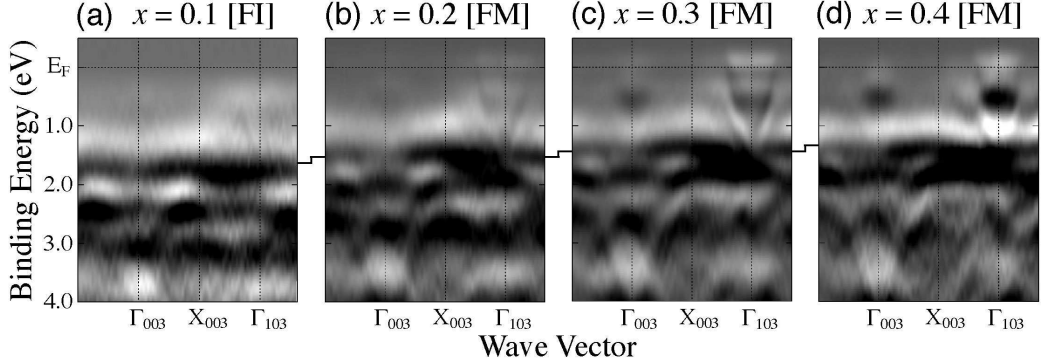}
\caption{Experimental band structure of the La$_{1-x}$Sr$_{x}$MnO$_{3}$ thin films along the $\Gamma$-$X$ direction [(a) $x = 0.1$, (b) $x = 0.2$, (c) $x = 0.3$, and (d) $x = 0.4$] obtained from the ARPES measurements.  Dark parts correspond to energy bands.  Solid lines represent the chemical potential shift determined by AIPES studies \cite{HoribaK:2005}.}
\label{fig1}
\end{center}
\end{figure}

Figure~1 shows the compositional dependence of the experimental band structure for LSMO thin films along the $\Gamma$-$X$ direction.  The band structure is visualized by plotting on gray scale the second derivative of the ARPES spectra, which were measured at a photon energy of 88 eV \cite{ChikamatsuA:2006}.  The ARPES spectra nearly reflect the band dispersion along the $\Gamma$-$X$ direction in the tetragonal Brillouin-zone of the epitaxial LSMO thin films because of the $k_{z}$-integrated density of states due to the effect of $k_{z}$-broadening \cite{ChikamatsuA:2006, WadatiH:2006}.  For ferromagnetic metallic LSMO $x = 0.4$ [Fig.~1(d)], we clearly found the existence of a parabolic conduction band near $E_{F}$ which forms an electron FS centered at the $\Gamma$ point \cite{ChikamatsuA:2006, ShiM:2004}.  In comparison with the band-structure calculation based on the local density approximation (LDA) + $U$ \cite{ChikamatsuA:2006}, the conduction band is ascribed to the Mn 3$de_{g}$ in-plane $3x^{2}-r^{2}$ states responsible for the half-metallic nature of LSMO.  Detailed analysis of the ARPES spectra has revealed that the size of the electron FS is consistent with the prediction of the band-structure calculation, while the width of the occupied part of the conduction band in experiment is significantly narrower than that in the calculation, indicative of the strong electron-electron correlation in LSMO \cite{ChikamatsuA:2006}.  On the other hand, several highly dispersive bands at binding energies higher than 2.0 eV mainly originate from the O $2p$ dominant states, while the dispersionless feature located at about 1.5 eV and the weakly dispersive features at 0.6 - 1.5 eV are assigned to the Mn $3d$ out-of-plane $y^{2}-z^{2}$ state and the Mn $3dt_{2g}$ states, respectively \cite{ChikamatsuA:2006}.

\begin{figure}
\begin{center}
\includegraphics[width=14cm]{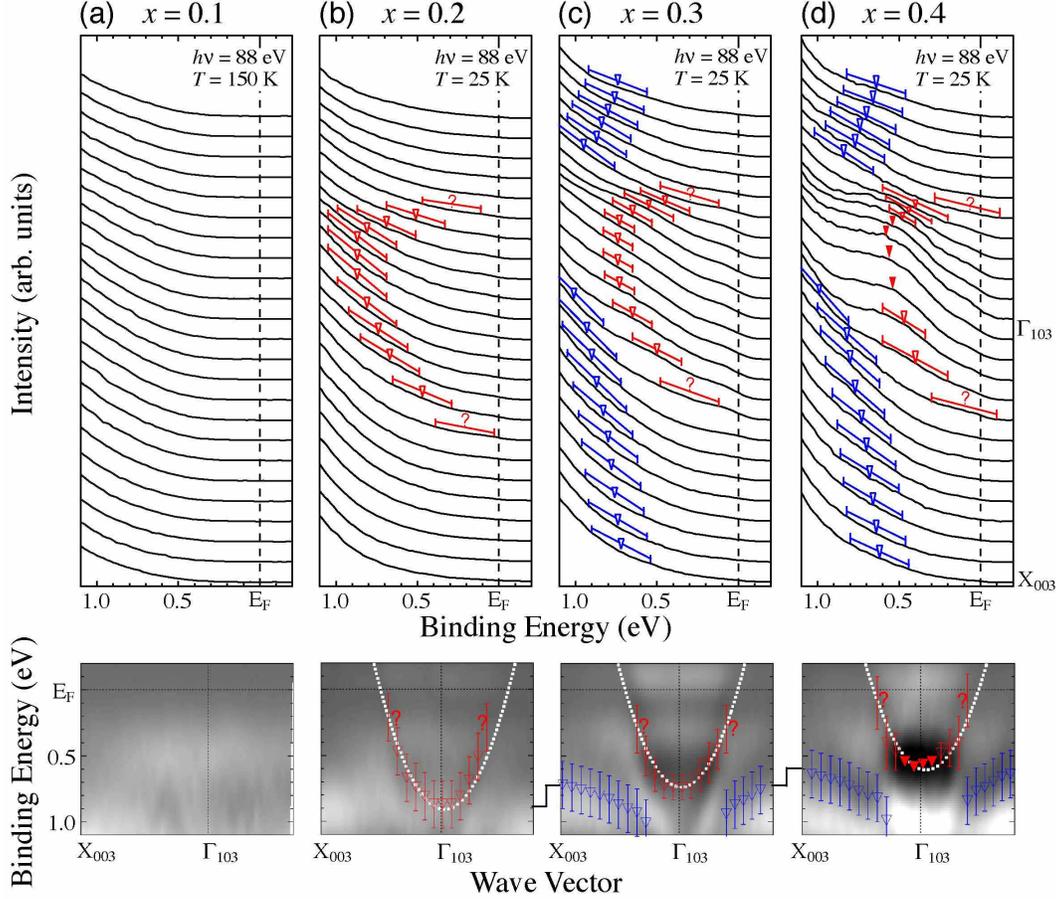}
\caption{(Color online) \textit{In-situ} ARPES spectra measured over a narrow energy range near $E_{F}$ around the $\Gamma_{103}$ point of La$_{1-x}$Sr$_{x}$MnO$_{3}$ thin films (upper panels) [(a) $x = 0.1$, (b) $x = 0.2$, (c) $x = 0.3$, and (d) $x = 0.4$].  The corresponding experimental band structure plots are shown in the lower panels.  Solid and open triangles and question marks represent the energy positions of prominent, medium, and weak structures, respectively.  The line segments indicate the upper and lower bounds for the dispersive features.  The white dashed line represents a fit of the energy positions of dispersive ARPES features to a parabola.}
\label{fig2}
\end{center}
\end{figure}

With decreasing hole concentration, the band structures at higher binding energies monotonically shift toward the higher binding energy side in a rigid-band manner.  The energy shift of these features from $x = 0.4$ to 0.1 is estimated to be 300 meV, which is in excellent agreement with the chemical potential shift determined by the angle-integrated (AI) PES \cite{HoribaK:2005} as well as core-level PES studies \cite{MatsunoJ:2002}.  The rigid-band like behavior is further confirmed by the fact that overall features of these bands do not exhibit essential changes upon hole doping.  These results suggest that the composition-dependent change in the electronic structures of LSMO is well described in the framework of the rigid-band model.

\begin{figure}
\begin{center}
\includegraphics[width=6cm]{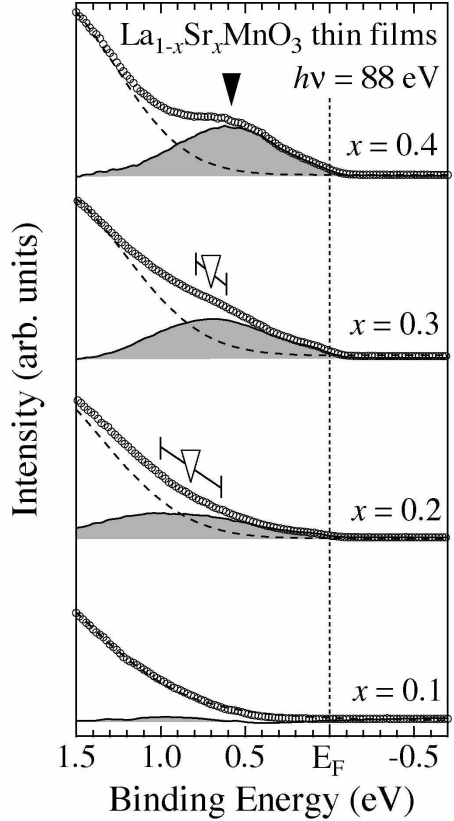}
\caption{\textit{In-situ} ARPES spectra near $E_{F}$ of La$_{1-x}$Sr$_{x}$MnO$_{3}$ thin films at the $\Gamma_{103}$ point (open circles) and dispersive ARPES features (solid curves) extracted by subtracting the contribution from the O $2p$ dominant states (dashed curves) from the original ARPES spectra.  Note that the perceptible spectral weight on the lower binding energy side for LSMO $x = 0.1$ is due to a thermal broadening effect.}
\label{fig3}
\end{center}
\end{figure}

From the rigid-band model, it is expected that the parabolic conduction band at the $\Gamma$ point should also show the same behavior.  However, in going from $x = 0.4$ to $x = 0.1$, the conduction band gradually smears out and finally disappears at $x = 0.1$, reflecting the MI transition in LSMO films at $x = 0.1$ - 0.2 \cite{HoribaK:2005}, in sharp contrast to that at higher binding energies.  Hereafter, we focus our attention on the observed anomalous behavior of the conduction band centered at the $\Gamma$ point.  In order to see the changes near $E_{F}$ in more detail, we show in Fig.~2 the ARPES spectra around the $\Gamma$ point in the near-$E_{F}$ region as a function of $x$, together with their experiment band structures on an expanded scale.  As expected from Fig.~1, the ARPES spectral weight corresponding to the parabolic conduction band gradually reduces with decreasing hole doping level from $x = 0.4$ to $x = 0.1$ and completely disappears at $x = 0.1$.  On the other hand, the bottom of the conduction band seems to monotonically shift toward higher binding energy with decreasing hole doping as if the energy position of the conduction band is dominated by the chemical potential shift of LSMO thin films.

Next we shall examine in more detail whether the ``remnant'' conduction band indeed exhibits the rigid band behavior.  In Fig.~3, we show the ARPES spectra at the $\Gamma$ point, where the dispersive ARPES feature corresponding to the bottom of the conduction band has been extracted by subtracting the spectral contribution of O $2p$ dominant states from the ARPES spectra.  The spectral contribution is simulated by a linear combination of Gaussian functions and an integral background \cite{HoribaK:2005}.  As shown in Fig.~3, the rigid shift for the peak at the $\Gamma$ point of about 200 meV from $x = 0.4$ to 0.2 is in excellent agreement with that predicted from the chemical potential shift of LSMO films, indicating that the energy position of the conduction band also obeys the rigid-band shift.  The rigid-band behavior of the conduction band is further confirmed in the band mapping of the second derivative ARPES spectra as shown in the bottom panel of Fig.~2, where we determined the energy positions in the same way as in Fig.~3 and fitted them to a parabola.  That is, the energy dispersions are well reproduced by the rigid shift of the parabola.  It should be noted that the estimated Fermi momentum for $x = 0.4$ by extrapolating the ARPES peak positions is in good agreement with that of band-structure calculation \cite{ChikamatsuA:2006}.  

Combined with previous AIPES results, the ARPES results give a much better understanding of how the spectral weight is redistributed and shifted near $E_{F}$ of LSMO.  While AIPES measurements showed the expected drop of spectral weight corresponding to the loss of electron from the $e_{g}$ states upon increased hole doping \cite{HoribaK:2005, SaitohT:1995}, ARPES spectra clearly show the increment of the spectral weight of the Mn $3de_{g}$-derived states.  These seeming contradictions could be reconciled by noticing a subtle energy difference of the $e_{g}$ feature in the AIPES and ARPES spectra and by considering the two components regarding the Mn $3de_{g}$ states, namely coherent and incoherent parts as schematically illustrated in Fig.~4: the parabolic conduction band in the ARPES spectra near $E_{F}$ corresponds to the coherent part, while the relatively broad AIPES spectra represent both the coherent and incoherent parts \cite{HoribaK:2005}.  The remnant behavior of the conduction band stems from the spectral weight transfer from the coherent band states to the incoherent states at 0.8 - 1.4 eV for $x = 0.4$ - 0.1, respectively \cite{HoribaK:2005}.  The absence of the incoherent part in the present ARPES spectra may be due to submergence in the tail of O $2p$ states located at 2 eV \cite{ChikamatsuA:2006}.  In fact, the reduction of spectral weight at $E_{F}$ with decreasing $x$ was clearly observed in the previous AIPES measurements \cite{HoribaK:2005}. 

\begin{figure}
\begin{center}
\includegraphics[width=14cm]{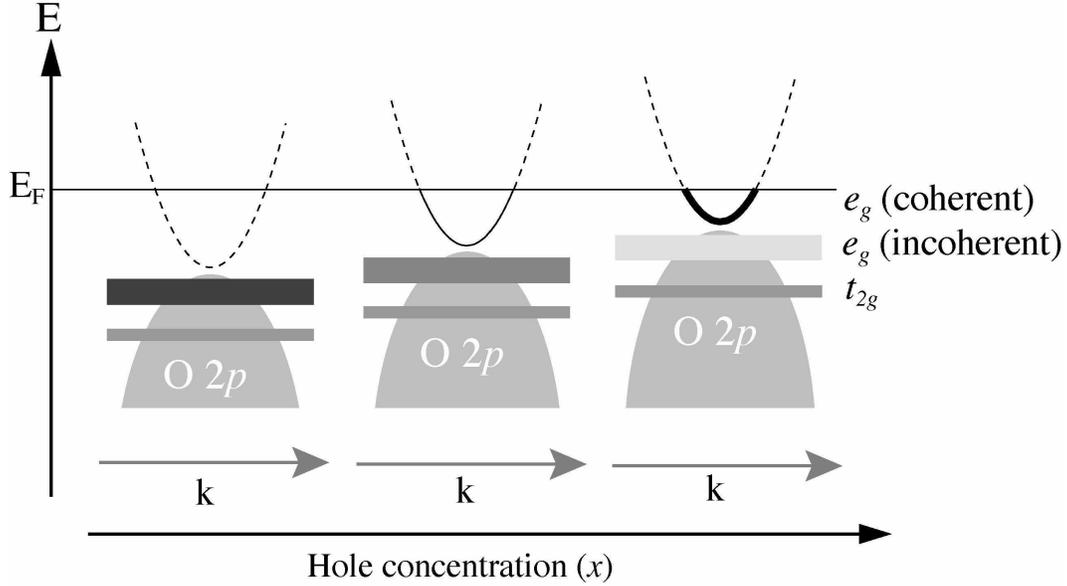}
\caption{A schematic illustration for the composition dependence of the electronic structure of La$_{1-x}$Sr$_{x}$MnO$_{3}$. }
\label{fig4}
\end{center}
\end{figure}

Next, we discuss the anomalous reduction of ARPES spectral weight near $E_{F}$ in samples with smaller $x$ and relate the loss to pseudogap formation.  The two characteristic features of manganite ARPES spectra are the vanishing or very small spectral weight at $E_{F}$, and the anomalously broad spectral features \cite{SunZ:2006, MannellaN:2005, ShiM:2004, ChikamatsuA:2006, DessauDS:1998}.  The origin of the low spectral weight at $E_{F}$, i.e. the pseudogap behavior in ARPES spectra is caused by strong electron-phonon coupling: when an electron is strongly coupled to collective excitations that can have various energies, a peak in an ARPES spectrum splits and the spectral shape is dominated by an envelope of many individual peaks.  In fact, as shown in Fig.~2, the ARPES spectra of metallic LSMO exhibit the pseudogap behavior near $E_{F}$.  The small spectral weight at $E_{F}$ and the anomalously broad incoherent spectral features seem to be common features in ARPES spectra of strongly correlated oxides.  For example, in the carrier doping evolution of ARPES spectra of lightly doped La$_{2-x}$Sr$_{x}$CuO$_{4}$ \cite{YoshidaT:2003} and Ca$_{2-x}$Na$_{x}$CuO$_{2}$Cl$_{2}$ \cite{ShenKM:2005, ShenKM:2004}, the spectral weight is redistributed toward the higher binding energy side with decreasing $x$, and can be described by the Franck-Condon broadening (FCB) scenario \cite{DessauDS:1998, ShenKM:2005, ShenKM:2004}.

Although the FCB scenario explains adequately  the characteristic ARPES features near $E_{F}$ of metallic LSMO films, the hole-doping evolution of LSMO films does not seem to be described by this scenario alone.  In the FCB scenario, the ARPES spectral weight (sum of spectral weight for a QP peak closest to $E_{F}$ and an envelope of many individual peaks) should be conserved within the energy range of electron-phonon couplings irrespective of carrier concentration \cite{ShenKM:2004}.  In LSMO, this energy range is evaluated to be 0 - 200 meV \cite{HartingerC:2004, MitraJ:2002}, in line with the present ARPES results.  However, in present results, the ARPES spectral weight itself is reduced and finally vanishes with decreasing $x$, indicating that spectral weight transfer toward higher binding energies other than the FCB mechanism must be considered.  The exotic physics in manganites as well as in other transition metal oxides is due to strong many-body interactions.  In the present system, both electron-electron and electron-phonon interactions must be considered.  Therefore, it is natural to conclude that the observed reduction of spectral weight is attributed to the spectral weight transfer from the coherent state in the near-$E_{F}$ region to the lower Hubbard band (incoherent states) with an energy separation of $U/2$, as explicitly predicted by the calculation based on dynamical mean field theory (DMFT) \cite{ZhangXY:1993, KajueterH:1996}.  In other words, there are two interactions with different energy scales that determine the spectral behavior of LSMO.  One is the electron-electron interaction, which dominates the spectral weight transfer between the coherent part and the incoherent part (Hubbard bands) on an energy scale of $U$, and the other is the electron-phonon interaction, which dominates spectral weight distribution in a narrow region near $E_{F}$, on the energy scale of multiple phonons.  In fact, recent calculation based on the $t$-$J$ model incorporating electron-phonon interaction on a single photohole in a Mott insulator has predicted the existence of the two different energy scales in theoretical ARPES spectra \cite{MishchenkoAS:2004}: the spectral weight transfer from the dispersive coherent states to the high-energy incoherent states due to the electron-electron interaction, and the broadening of the coherent states derived from the spectral redistribution due to the electron-phonon interaction.  For a better understanding of the anomalous spectral transfer in manganites with changing $x$, it is important to take into account both interactions appropriately in theoretical considerations.  More systematic and detailed ARPES studies on LSMO as well as on other manganites as a function of carrier concentration are desired.

In conclusion, we have performed an \textit{in situ} ARPES study on LSMO thin films to investigate the evolution of electronic structure as a function of hole concentration $x$.  The spectral weight of the conduction band in the metallic phase gradually smears out with decreasing hole concentration and almost disappears in the insulating phase, whereas the entire band structure, including the energy position of the conduction band, monotonically shifts in a rigid-band manner.  These results are not described within the framework of the one electron picture and indicate that spectral weight transfer from the coherent band to the incoherent states (lower Hubbard band) dominates the gap formation in LSMO associated with the filling-control MI transition.  The pseudogap behavior also exists in the ARPES spectra near $E_{F}$ in the metallic LSMO irrespective of $x$, suggesting that the novel electronic structure of manganites is dominated by two different interactions with different energy scale: electron-electron interaction and electron-phonon interaction.

This work was supported by Grants-in-Aid for Scientific Research (S17101004 and A16204024) from JSPS.

\bibliography{LSMO_ARPES_Com}

\end{document}